\documentclass[conf]{new-aiaa}
\usepackage[utf8]{inputenc}

\usepackage{graphicx}
\usepackage{amsmath}
\usepackage[version=4]{mhchem}
\usepackage{siunitx}
\usepackage{longtable,tabularx}
\usepackage{caption}
\usepackage{subcaption}
\usepackage{nomencl}
\makenomenclature

\usepackage{soul} 
\usepackage{xcolor}

\newcommand{\figref}[1]{Fig.~\ref{#1}}
\newcommand{\tabref}[1]{Table~\ref{#1}}

\newcommand{\secref}[1]{Section~\ref{#1}}

\newcommand{\obs}[0]{\boldsymbol{o}}
\newcommand{\Obs}[0]{O}
\newcommand{\ObsFunc}[0]{\Omega}
\newcommand{\state}[0]{\boldsymbol{x}}

\newcommand{\action}[0]{\boldsymbol{u}}

\newcommand{\control}[0]{u}
\newcommand{\controlmax}[0]{\control_{\rm max}}

\newcommand{\reward}[0]{r}

\newcommand{\policy}[0]{\pi}

\newcommand{\Reward}[0]{\mathbf{R}}

\newcommand{\sunangle}[0]{\theta_{\rm Sun}}
\newcommand{\sunangledot}[0]{\dot{\theta}_{\rm Sun}}
\newcommand{\numpoints}[0]{n_p}
\newcommand{\velocity}[0]{\boldsymbol{v}}

\newcommand{\position}[0]{\boldsymbol{p}}
\newcommand{\deltav}[0]{\Delta V}
\newcommand{\deltat}[0]{\Delta t}

\newcommand{\udes}[0]{\boldsymbol{u}_{\rm des}}
\newcommand{\uact}[0]{\boldsymbol{u}_{\rm act}}
\newcommand{\meanmotion}[0]{\eta}
\newcommand{\mass}[0]{m_{\rm d}}

\title{Demonstrating Reinforcement Learning and Run Time Assurance for Spacecraft Inspection Using Unmanned Aerial Vehicles}

\author{
Kyle Dunlap\footnote{AI Software Developer, Intelligent Systems Division, AIAA Professional Member.}\textsuperscript{1} and 
Nathaniel Hamilton\footnote{AI Scientist, Intelligent Systems Division.}\textsuperscript{1}
}
\affil{Parallax Advanced Research, Beavercreek, OH, 45431, USA}

\author{
Zachary Lippay\footnote{Team Lead, Dynamics and Controls, AIAA Professional Member.} and
Matthew Shubert\footnote{Senior Robotics Engineer, Dynamics and Controls, AIAA Professional Member.}
}
\affil{Verus Research, Albuquerque, NM, 87123, USA}

\author{
Sean Phillips\footnote{Technology Advisor, Space Control Branch, AIAA Professional Member.}
}
\affil{Air Force Research Laboratory, Kirtland Air Force Base, NM, 87117, USA}

\author{
Kerianne L. Hobbs\footnote{Safe Autonomy Lead, Autonomy Capability Team (ACT3), AIAA Associate Fellow.} 
}
\affil{Air Force Research Laboratory, Wright-Patterson Air Force Base, OH, 45433, USA}

\begin{document}

\maketitle
\begingroup\renewcommand\thefootnote{1}
\begin{NoHyper}
\footnotetext{These authors contributed equally to this work.}
\end{NoHyper}
\endgroup

\begin{abstract}

On-orbit spacecraft inspection is an important capability for enabling servicing and manufacturing missions and extending the life of spacecraft. However, as space operations become increasingly more common and complex, autonomous control methods are needed to reduce the burden on operators to individually monitor each mission. In order for autonomous control methods to be used in space, they must exhibit safe behavior that demonstrates robustness to real world disturbances and uncertainty.
In this paper, neural network controllers (NNCs) trained with reinforcement learning are used to solve an inspection task, which is a foundational capability for servicing missions. Run time assurance (RTA) is used to assure safety of the NNC in real time, enforcing several different constraints on position and velocity. The NNC and RTA are tested in the real world using unmanned aerial vehicles designed to emulate spacecraft dynamics. The results show this emulation is a useful demonstration of the capability of the NNC and RTA, and the algorithms demonstrate robustness to real world disturbances.
\end{abstract}

\section{Introduction}


\lettrine{A}{s} the number of spacecraft in orbit continues to grow, it is becoming more difficult for human operators to monitor and plan for each mission. This motivates the need for autonomous spacecraft operations, which have the capability to scale to complex missions without the need for constant operator intervention. Specifically, \textit{On-orbit Servicing, Assembly, and Manufacturing} (OSAM) missions are critical for sustained space operation, where the life of spacecraft can be extended. However, these missions require multiple spacecraft to operate within close proximity to one another, where safety is critical. As a result, operators must have a high level of trust in the autonomous control systems if autonomy will ever be used in the real world \cite{10115976}.

One method for autonomous control is \textit{Reinforcement Learning} (RL), which is a fast growing field in machine learning that has shown great success across many complex tasks such as Go \cite{silver2016mastering} and Starcraft \cite{starcraft2019}. Deep RL uses a \textit{Neural Network Controller} (NNC) to autonomously control the system, which can occur at the guidance, navigation, or control level. Deep RL has shown success across many simulated space tasks, including inspection \cite{LeiGNC22, AurandGNC23, Brandonisio2021}, docking \cite{oestreich2021autonomous, broida2019spacecraft,copp2023Learning}, proximity operations \cite{hovell2021deep}, and long duration missions \cite{stephenson2024using, harris2022generation}.

However, due to the trial and error nature of RL and the difficulty of verifying the behavior of an NNC, safety assurance methods are a high priority for building trust. One method of assuring safety is \textit{Run Time Assurance} (RTA), which filters the output of the NNC to ensure the control is safe. Specifically, this paper considers the use of \textit{Control Barrier Functions} (CBFs) to assure safety. Using CBFs is a popular method for assuring safety in complex systems such as autonomous driving \cite{ames2016control} and fixed-wing aircraft \cite{squires2022composition}, as well as many space related tasks including docking \cite{breeden2021guaranteed}, rendezvous \cite{agrawal2021safe}, and proximity operations \cite{mote2021natural, mcquinn2024run}. RTA has also been used to assure safety during RL training for simulated spacecraft docking \cite{hamilton2023ablation, dunlap2023rta_rl}.

While all of these methods have shown success in simulation, they must be tested on real world systems to gain the trust of operators. Due to the high cost of spacecraft and complexity of the space environment, testing these algorithms on orbit is expensive and risky. Instead, this paper focuses on testing in a laboratory where unmanned aerial vehicles are used to simulate spacecraft dynamics. Deploying the trained NNC on a real-world platform is known as a \textit{sim2real} transfer, which can be challenging as the NNC often performs differently than expected in the real world \cite{hamilton2022zero}. By demonstrating the NNC and RTA in the real world, this allows operators to visualize the expected behavior of the system and understand how it will perform in unexpected situations.

The main contributions of this work are: 1) successfully testing an NNC and RTA filter on a real world system (quadrotors that mimic space dynamics) and 2) using real world state data to test the robustness of the NNC and RTA. This paper builds on previous work to use RL to train an agent to complete the inspection problem \cite{vanWijkAAS_23}, and developing safety constraints for inspection that can be enforced by RTA \cite{dunlap2023run}.

The remainder of this work is organized as follows. Section~\ref{sec:background} provided background information about reinforcement learning, run-time assurance, and the laboratory environment. Then in Section~\ref{sec:exp_setup}, we the experimental setup including the satellite mission and the training policy setup. In Section~\ref{sec:results}, we provide the main results from this study and provide the concluding remarks in Section~\ref{sec:conclusion}. 

\section{Background}\label{sec:background}

This section provides an introduction for reinforcement learning, run time assurance, and the real world testing laboratory.

\subsection{Reinforcement Learning}

\begin{figure}[htbp]
    \centering
    \includegraphics[width=0.6\columnwidth]{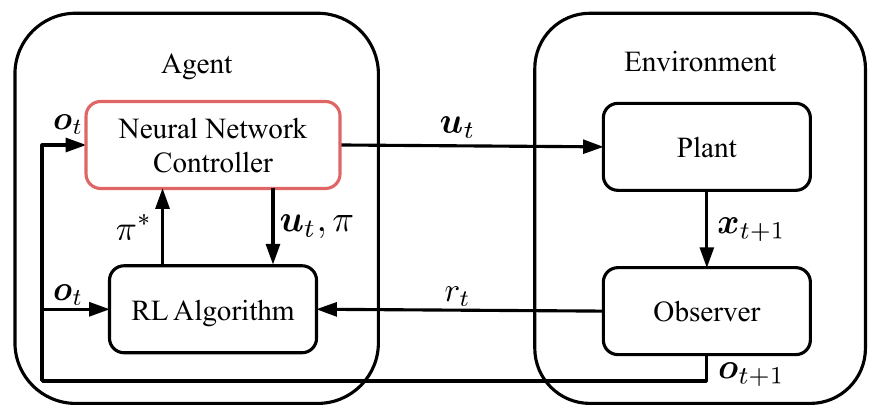} 
    \caption{Basic formulation of Deep Reinforcement Learning for a control problem.}
    \label{fig:drl_basic}
\end{figure}

\emph{Reinforcement Learning} is a form of machine learning in which an agent acts in an environment, learns through experience, and improves the performance of the learned behavior function, i.e. policy $\policy$, based on the observed reward. In \emph{Deep Reinforcement Learning} (RL) the policy is approximated using a deep neural network, which becomes the \emph{Neural Network Controller} (NNC) after training is completed\footnote{This work considers zero-shot policy transfer for using the trained NNC in the real world. That means there is no additional training done in the real world. The final policy produced by the RL training is used without modification in the real world experiments.}. The basic construction of the RL training approach for a control problem is shown in \figref{fig:drl_basic}. The agent consists of the NNC and RL algorithm, and the environment consists of a plant and observer model. The environment can be comprised of any dynamical system, from Atari simulations (\cite{hamilton2020sonic, alshiekh2018safe}) to complex robotics scenarios (\cite{brockman2016openai, fisac2018general, henderson2018deep, mania2018simple, jang2019simulation, bernini2021few}). 

Reinforcement learning is based on the \textit{reward hypothesis} that all goals can be described by the maximization of expected cumulative reward \cite{silver2015}. During training, the agent chooses an action, $\action_t$, based on the input observation, $\obs$. The action is then executed in the environment, transitioning the environment to the next state, $\state_{t+1}$, according to the plant model. However, instead of receiving the state information, the agent receives an observation $\obs_{t+1} \in \Obs$  which depends on the new state of the environment, $\state_{t+1}$, and the recent action, $\action_t$, with probability $\ObsFunc(\obs_{t+1} \mid \state_{t+1}, \action_t)$. Finally, the agent receives a reward $\reward_t$ according to the reward function $\Reward(\state_t, \action_t, \state_{t+1})$.

The process of executing an action and receiving a reward and next observation is referred to as a \emph{timestep}. Relevant values, like the input observation, action, and reward are collected as a data tuple, i.e. \emph{sample}, by the RL algorithm to update the current NNC policy, $\policy$, to an improved policy, $\policy^*$. How often these updates are done is dependent on the RL algorithm.

This work focuses on NNCs trained using the Proximal Policy Optimization (PPO) algorithm. PPO was first introduced in \cite{schulman2017proximal}. The  ``proximal'' in PPO refers to how the algorithm focuses on iteratively improving the policy in small increments to prevent making large changes in the policy that lead to drops in performance. PPO was selected because of its widespread use in a variety of RL tasks and previous demonstrated success in the space domain \cite{broida2019spacecraft, oestreich2021autonomous, hamilton2023ablation, dunlap2023rta_rl, vanWijkAAS_23}.

\subsection{Run Time Assurance}

\begin{figure}
    \centering
    \includegraphics[width=.8\columnwidth]{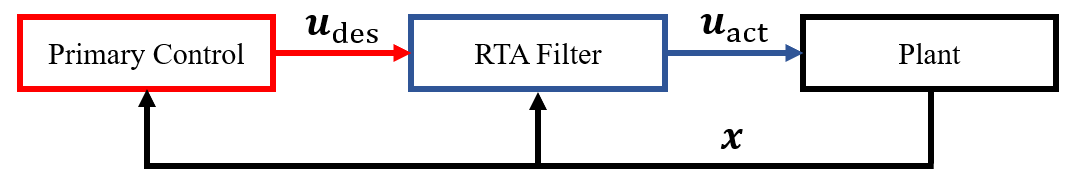}
    \caption{Feedback control system with RTA filter. Components with low safety confidence are outlined in red, and components with high safety confidence are outlined in blue.}
    \label{fig:RTA_Filter}
\end{figure}

Run Time Assurance (RTA) is an online safety assurance technique that filters an unverified primary controller to assure safety of the system \cite{hobbs2023runtime}. The primary controller can be a human operator, an NNC trained using RL, or any other control techniques. \figref{fig:RTA_Filter} shows a feedback control system with RTA, where the RTA is completely separate from the primary controller. This allows the primary controller to focus on completing the task, while the RTA filter focuses on assuring safety.

For the analyses in this paper, safety is enforced using Control Barrier Functions (CBFs) \cite{ames2019control}. First, consider a continuous time control affine dynamical system defined as,
\begin{equation} \label{eq:fxgu}
   \boldsymbol{\dot{x}} = f(\boldsymbol{x}) + g(\boldsymbol{x})\boldsymbol{u},
\end{equation}
where $f:\mathcal{X} \rightarrow \mathbb{R}^n$ and $g:\mathcal{X} \rightarrow \mathbb{R}^{n \times m}$ are locally Lipschitz continuous functions. $\boldsymbol{x} \in \mathcal{X} \subseteq \mathbb{R}^n$ denotes the state vector, $\mathcal{X}$ is the set of all possible state values, $\boldsymbol{u}\in \mathcal{U} \subseteq\mathbb{R}^m$ denotes the control vector, and $\mathcal{U}$ is the admissible control set.


For this system, safety can be defined through a continuously differentiable function $h:\mathcal{X} \rightarrow \mathbb{R}$. The superlevel set of this function is defined as $\mathcal{C}_S = \{\boldsymbol{x} \in \mathcal{X} : h(\boldsymbol{x}) \ge 0\}$, where $\mathcal{C}_S$ is referred to as the safe set. If $\mathcal{C}_S$ is forward invariant, where for every $\boldsymbol{x} \in \mathcal{C}_S$, $\boldsymbol{x}(t) \in \mathcal{C}_S, \, \forall t \geq 0$, then the system cannot leave $\mathcal{C}_S$ and it will always be safe. 
Nagumo's condition \cite{nagumo1942lage} is used to enforce this condition and ensure $\dot{h}(\boldsymbol{x}) \geq 0$ along the boundary of $\mathcal{C}_{\rm S}$. An extended class $\mathcal{K}$ function $\alpha : \mathbb{R} \rightarrow \mathbb{R}$ is used to relax the condition away from the boundary, where $\alpha$ is strictly increasing and has the property $\alpha(0) = 0$. 
As a result, $h$ is a CBF if there exists an extended class $\mathcal{K}$ function $\alpha$ and control $\boldsymbol{u} \in \mathcal{U}$ such that,
\begin{equation}\label{eq:cbf_condition}
    \sup_{\boldsymbol{u} \in \mathcal{U}} [L_fh(\boldsymbol{x}) + L_gh(\boldsymbol{x}) \boldsymbol{u} ] \geq - \alpha(h(\boldsymbol{x})), \mkern9mu \forall \boldsymbol{x} \in \mathcal{C}_S,
\end{equation}
where $L_f$ and $L_g$ are Lie derivatives of $h$ along $f$ and $g$ respectively. This type of CBF is known as a zeroing CBF.

This paper uses the RTA technique known as the Active Set Invariance Filter (ASIF) \cite{ASIF_2018}, which uses CBFs to filter the primary controller and assure safety of the system. ASIF is advantageous as it is minimally invasive towards the primary controller and has the capability to enforce multiple safety constraints simultaneously. ASIF is formulated with a Quadratic Program (QP), where the objective function is the $l^2$ norm difference between the primary control $\udes$ and safe control $\uact$. This ensures the safe control will be as close as possible to the desired control. The CBFs are added to the QP as inequality constraints, where multiple CBFs can be enforced simultaneously provided they jointly render $\mathcal{C}_S$ forward invariant. For $N$ CBFs, the ASIF algorithm is defined as,

\begin{samepage}
\noindent \rule{1\columnwidth}{0.7pt}
\noindent \textbf{Active Set Invariance Filter}
\begin{equation}
\begin{gathered}
\boldsymbol{u}_{\rm act}(\boldsymbol{x}, \boldsymbol{u}_{\rm des})= \underset{\boldsymbol{u} \in \mathcal{U}}{\text{argmin}} \left\Vert \boldsymbol{u}_{\rm des}-\boldsymbol{u}\right\Vert\\
\text{s.t.} \mkern9mu L_f h(\boldsymbol{x}) + L_g h(\boldsymbol{x}) \boldsymbol{u} + \alpha(h(\boldsymbol{x})) \geq 0, \mkern9mu \forall i \in \{1,...,N\}
\end{gathered}\label{eq:optimization}
\end{equation}
\noindent \rule[7pt]{1\columnwidth}{0.7pt}
\end{samepage}

\subsection{LINCS}


The Local Intelligent Networked Collaborative Satellites (LINCS) laboratory is a simulation and emulation environment designed for testing multi-vehicle satellite autonomy, guidance, navigation, and control algorithms \cite{phillips2024emulation}. The lab uses quadrotors to emulate spacecraft motion inside a caged aviary, which represents the Hill's frame (i.e., a relative spacecraft frame or radial in-track cross-track (RIC) frame).
The guidance, navigation, and control algorithms are simulated using \textit{2-body dynamics with J2 gravitational perturbations}, and these trajectories are emulated using the quadrotors in the physical lab space such that they appear to follow the space trajectories. Importantly, these trajectories must be scaled from the space frame to the lab frame so they fit within the dimensions of the lab space. A Vicon motion capture system is used to determine the physical state of the quadrotors.

To assure safety in the lab, an additional lab RTA filter is applied to the control of each quadrotor. This RTA is based on quadrotor dynamics, rather than space dynamics, and is designed to prevent the quadrotors from colliding with each other or the walls of the lab, as well as to limit the acceleration along each axis. This allows unverified algorithms to be tested in the lab without the risk of damaging the quadrotors or the lab space. Further descriptions of the LINCS lab and quadrotor RTA algorithms are beyond the scope of this paper, but are provided in \cite{phillips2024emulation}.

\begin{figure}[htb]
\begin{subfigure}{0.5\textwidth}
\includegraphics[width=0.99\linewidth]{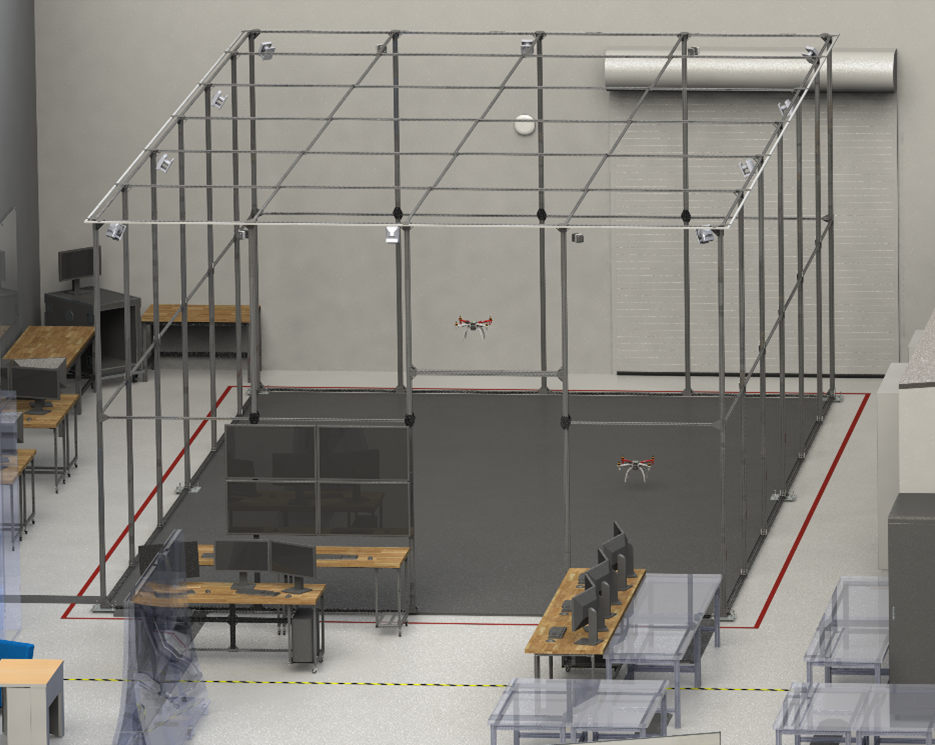} 
\label{fig:Aviary_CAD}
\caption{CAD model of the Aviary testbed facility.}
\end{subfigure}
\begin{subfigure}{0.5\textwidth}
\includegraphics[width=0.99\linewidth]{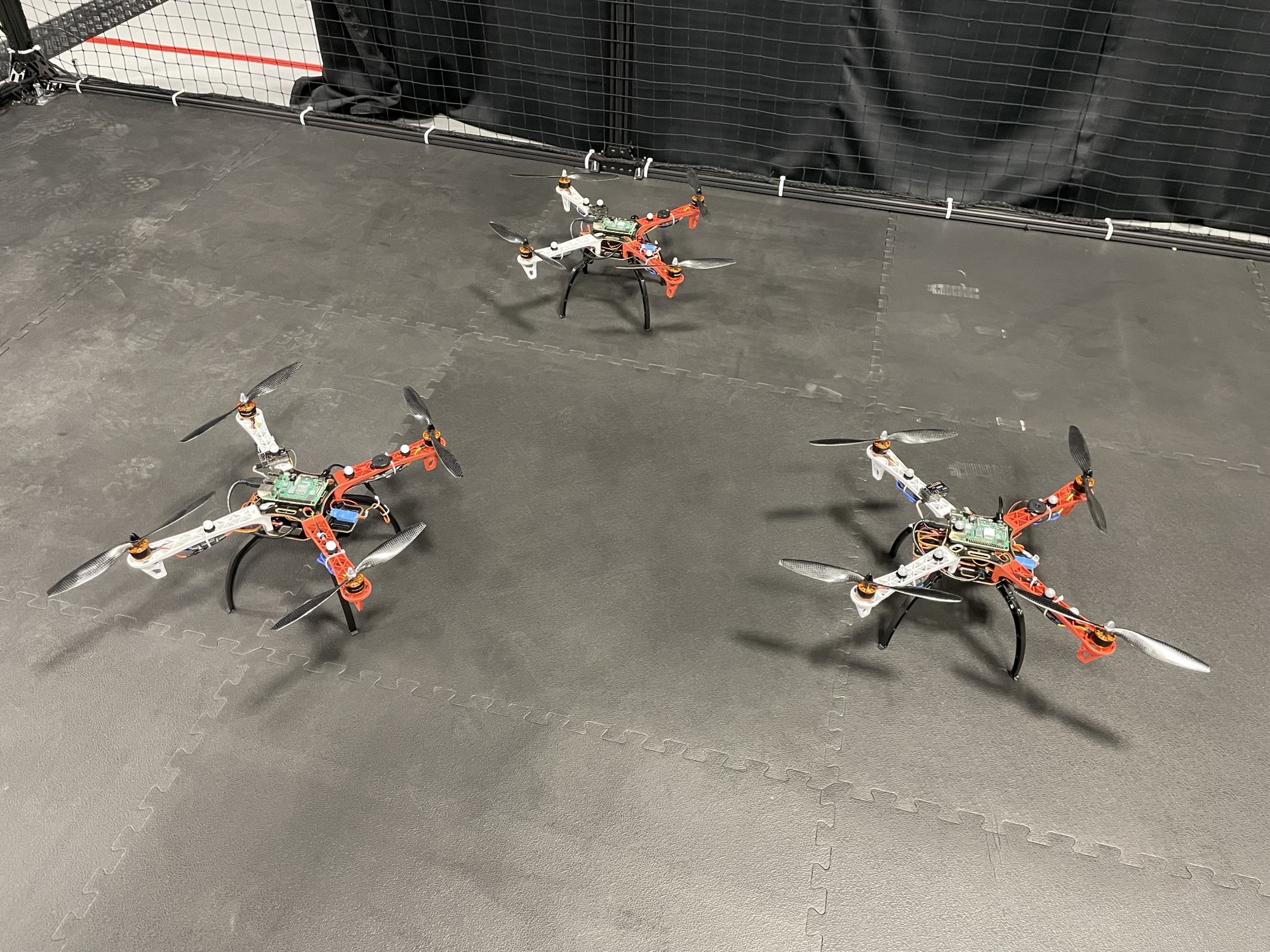}
\label{fig:Aviary_actual}
\caption{Squadron of quadcopters currently utilized in the lab.}
\end{subfigure}
\caption{Local Intelligent Network of Collaborative Satellites (LINCS) Laboratory overview}
\label{fig:Aviary}
\end{figure}

\section{Experimental Setup} \label{sec:exp_setup}

This section introduces the spacecraft inspection task, the associated RL training setup and RTA constraints for the task, and a description of all the experiments that were run in the lab.

\subsection{Inspection Task}

\begin{figure}[hbt]
    \centering
    \includegraphics[width=.5\columnwidth]{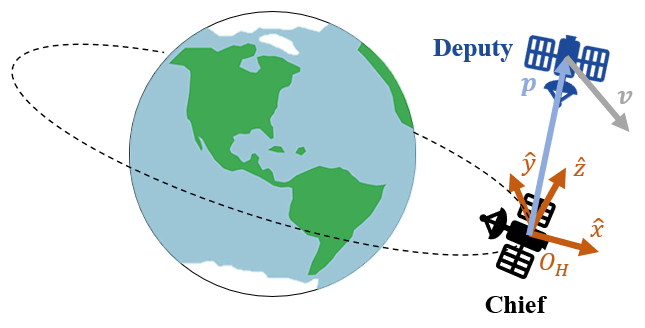}
    \caption{Deputy spacecraft in relation to a chief spacecraft in Hill's Frame.}
    \label{fig:Hills}
\end{figure}

For the spacecraft inspection task, an active ``deputy'' spacecraft attempts to navigate around and inspect a passive ``chief'' spacecraft \cite{vanWijkAAS_23}. As shown in \figref{fig:Hills}, the chief is assumed to be in a circular orbit around the Earth, and motion of the deputy spacecraft is described relative to the chief in Hill's frame \cite{hill1878researches}. Here, the origin $\mathcal{O}_H$ is located at the chief's center of mass, the unit vector $\hat{x}$ points away from the center of the Earth, $\hat{y}$ points in the direction of motion of the chief, and $\hat{z}$ is normal to $\hat{x}$ and $\hat{y}$. A sphere of 99 uniformly distributed points with a radius of $10$~m is used to represent the chief. The deputy can inspect these points if they are within its field of view, and in some cases if the points are illuminated by the Sun. It is assumed that the deputy always points towards the chief, as the attitude is not modeled. It is also assumed that the Sun remains at a fixed position in relation to the Earth, where it appears to rotate co-planar with the chief's orbit at a constant rate in Hill's frame. This simplification allows the position of the Sun to be represented by one angle.

\subsubsection{Dynamics}

To train the NNC and develop RTA constraints, the linearized Clohessy-Wiltshire equations \cite{clohessy1960terminal} are used to model the dynamics of the deputy spacecraft. 
%
The linearized relative motion dynamics of the deputy are defined as, 
\begin{equation} \label{eq: system dynamics}
    \dot{\state} = A {\state} + B\action,
\end{equation}
where $\state$ is the state vector $\state=[x,y,z,\dot{x},\dot{y},\dot{z}]^T \in \mathbb{R}^6$, $\action$ is the control vector, $\action= [F_x,F_y,F_z]^T \in [-\controlmax, \controlmax]^3$, and,
\begin{align}
\centering
    A = 
\begin{bmatrix} 
0 & 0 & 0 & 1 & 0 & 0 \\
0 & 0 & 0 & 0 & 1 & 0 \\
0 & 0 & 0 & 0 & 0 & 1 \\
3\meanmotion^2 & 0 & 0 & 0 & 2\meanmotion & 0 \\
0 & 0 & 0 & -2\meanmotion & 0 & 0 \\
0 & 0 & -\meanmotion^2 & 0 & 0 & 0 \\
\end{bmatrix}, 
    B = 
\begin{bmatrix} 
 0 & 0 & 0 \\
 0 & 0 & 0 \\
 0 & 0 & 0 \\
\frac{1}{\mass} & 0 & 0 \\
0 & \frac{1}{\mass} & 0 \\
0 & 0 & \frac{1}{\mass} \\
\end{bmatrix}.
\end{align}
Here, $\meanmotion=0.001027$~rad/s is the mean motion of the chief's orbit, $\mass=12$~kg is the mass of the deputy, $F$ is the force exerted by the deputy's thrusters along each axis, and $\controlmax=1$N is the maximum force.
The Sun appears to rotate in the $\hat{x}-\hat{y}$ plane in Hill's frame, where the unit vector pointing from the center of the chief spacecraft to the Sun, $\hat{r}_{S}$, is defined as,
\begin{equation}
    \hat{r}_{S} = [\cos{\sunangle}, \sin{\sunangle}, 0],
\end{equation}
where $\sunangledot=-\meanmotion$. It is assumed that the Sun is the only light source and nothing blocks the chief from being illuminated.

\subsubsection{RL Environment Setup}
\label{sec:RL_setup}

The main objective of the inspection task is to inspect all points on the chief spacecraft. However, a secondary objective is to minimize fuel use, which is considered in terms of $\deltav$,
\begin{equation}
    \deltav = \frac{|F_{x}| + |F_{y}| + |F_{z}|}{\mass} \deltat.
\end{equation}

Several different observations are available to the agent, where the environment is partially observable. 
First, the state information of position $\position = [x,y,z]$, velocity $\velocity = [\dot{x},\dot{y},\dot{z}]$, and sun angle $\sunangle$ are provided, where each position value is divided by $100$ (assuming the inspection task takes place within 100~m of the chief), and each velocity value is multiplied by $2$ such that common values fall within the range $[-1, 1]$. Related to the inspection points, two components are provided: the total number of inspected points during the episode $\numpoints$, which is divided by 100, and a unit vector pointing towards the nearest cluster of uninspected points $\hat{r}_{UPS}=[x_{UPS}, y_{UPS}, z_{UPS}]$ as determined by k-means clustering. Two different observation space configurations were trained for this environment.
The first configuration is referred to as ``no sensors'' and is defined as,
\begin{equation}
    \obs_{\rm no} = [x, y, z, \dot{x}, \dot{y}, \dot{z}].
\end{equation}
The second configuration is referred to as ``all sensors'' and is defined as,
\begin{equation}
    \obs_{\rm all} = [x, y, z, \dot{x}, \dot{y}, \dot{z}, \numpoints, \sunangle, x_{UPS}, y_{UPS}, z_{UPS}].
\end{equation}

\subsubsection{Safety Constraints}

For the inspection task, the following CBFs are used with an ASIF RTA filter to assure safety. Note that these constraints are not enforced during RL training. Further details and derivations for all constraints can be found in \cite{dunlap2023run}. 

First, the deputy shall avoid colliding with the chief. This CBF is defined as,
\begin{equation}
    h_1(\boldsymbol{x}) := \sqrt{2 a_{\rm max} [\Vert \boldsymbol{p} \Vert_2 - (r_{\rm d}+r_{\rm c})]} + \boldsymbol{v}_{{prc}} \geq 0,
\end{equation}
where $a_{\rm max}=0.078$~m/s$^2$ is the maximum acceleration of the deputy due to control and natural motion, $\boldsymbol{p}$ is the deputy's position, $r_{\rm d}=5$~m is the radius of the deputy, $r_{\rm c}=5$~m is the radius of the chief, and $\boldsymbol{v}_{{prc}}$ is the projection of the deputy's velocity in the direction of the chief. This constraint ensures that the deputy has enough time to slow down to avoid a future collision.

Second, the deputy shall not travel too far from the chief, allowing it to remain on task. This CBF is defined as,
\begin{equation}
    h_2(\boldsymbol{x}) := \sqrt{2 a_{\rm max} (r_{\rm max} - \Vert \boldsymbol{p} \Vert_2)} - \boldsymbol{v}_{{prr}} \geq 0,
\end{equation}
where $r_{\rm max}=1000$~m is the maximum radial distance from the chief, and $\boldsymbol{v}_{{prr}}$ is the projection of the deputy's velocity in the direction of the keep in zone. This constraint similarly ensures that the deputy has enough time to slow down to avoid exceeding the maximum distance.

Third, the deputy shall slow its velocity as it approaches the chief. This reduces the risk of a potential collision in uncertain environments. This CBF is defined as,
\begin{equation}
    h_3(\boldsymbol{x}) := \nu_0 + \nu_1\Vert \boldsymbol{p} \Vert_2 - \Vert \boldsymbol{v} \Vert_2 \geq 0,
\end{equation}
where $\nu_0 = 0.2$~m/s is a minimum allowable speed at the origin of Hill's frame, $\nu_1 = 2\meanmotion$~rad/s is a constant rate at which $\boldsymbol{p}$ shall decrease, and $\boldsymbol{v}$ is the deputy's velocity.

Fourth, the deputy shall not maneuver aggressively with large velocities. This is defined in terms of three separate CBFs,
\begin{equation}
    h_4(\boldsymbol{x}) := v_{\rm max}^2 - \dot{x}^2\geq 0, \quad h_5(\boldsymbol{x}) := v_{\rm max}^2 - \dot{y}^2\geq 0, \quad h_6(\boldsymbol{x}) := v_{\rm max}^2 - \dot{z}^2\geq 0,
\end{equation}
where $v_{\rm max}=1$~m/s is the maximum allowable velocity.

Safety for the system is defined as $\mathcal{C}_S = \{\boldsymbol{x} \in \mathcal{X} : h_i(\boldsymbol{x}) \ge 0\, , \, \forall i \in \{1, ..., 6\} \}$, where all CBFs are enforced simultaneously by the ASIF RTA.

\subsection{Training the Neural Network Controllers}

\begin{figure}
    \centering
    \includegraphics[width=0.8\columnwidth]{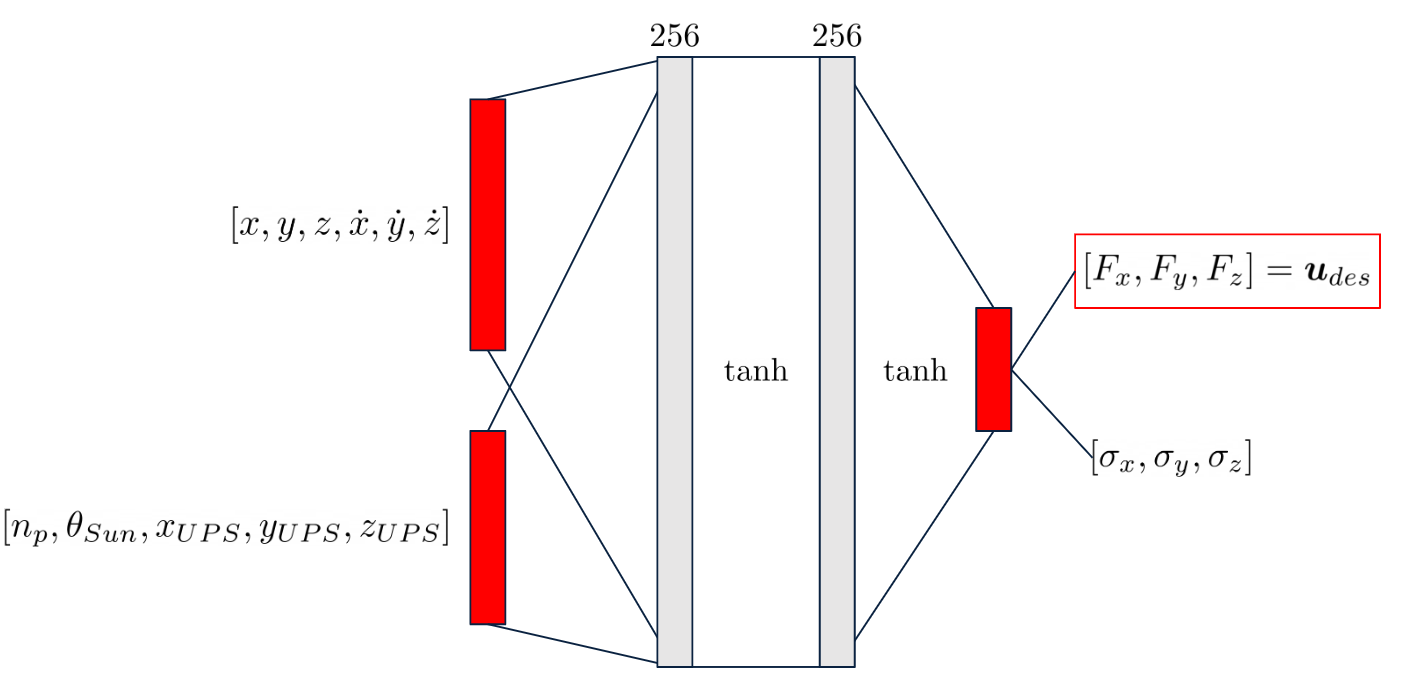}
    \caption{The NNC architecture for the ``all sensors'' variation of the inspection task. 
    }
    \label{fig:nnc_architecture}
\end{figure}

\begin{table}[htb]
    \centering
    \caption{PPO Training Hyperparameters}
    \begin{tabular}{ll} \hline
        Parameter                       & Value \\ \hline
        Number of SGD iterations        & 30 \\ 
        Discount factor $\gamma$        & 0.99 \\
        GAE-$\lambda$                   & 0.928544 \\
        Max episode length              & 1223 \\
        Rollout fragment length         & 1500 \\
        Train batch size                & 1500 \\
        SGD minibatch size              & 1500 \\
        Total timesteps                 & $5 \times 10^6$ \\
        Learning rate                   & $5 \times 10^{-5}$ \\
        KL initial coefficient          & 0.2 \\
        KL target value                 & 0.01 \\
        Value function loss coefficient & 1.0 \\
        Entropy coefficient             & 0.0 \\
        Clip parameter                  & 0.3 \\
        Value function clip parameter   & 10.0 \\
        \hline
    \end{tabular}
    \label{tab:Hyperparameters}
\end{table}

This work focuses on testing four NNCs trained to solve the two variations of the inspection task described in \secref{sec:RL_setup}. In both variations, the agent is learning to maximize performance according to the reward function,
\begin{equation}
    \reward = 0.1(\numpoints) - 0.1(\deltav),
\end{equation}
where $\numpoints$ is the total number of points inspected (maximum of 99), and $\deltav$ corresponds to the total amount of fuel used to complete the inspection. During RL training, $\deltat=10$ seconds. The PPO hyperparameters used to train the NNCs for these experiments are listed in \tabref{tab:Hyperparameters}. The NNC architecture for the ``all sensors'' variation is shown in \figref{fig:nnc_architecture}. The ``no sensors'' variation has the same architecture, but with the bottom input block removed.

The NNC architecture is a fully-connected Multi-Layer Perceptron (MLP) network with two hidden layers with 256 nodes each. The hidden layers have a $\tanh$ activation function\footnote{This is the common/default architecture used for PPO.}. The NNC has either 6 or 11 inputs, depending on the observation space used. The output layer consists of 6 nodes, which represent the mean and variance for each control value in $\action$ for the stochastic policy used during training. In these experiments, the output is always the mean value as show by the red box in \figref{fig:nnc_architecture}. The inputs are normalized such that typical values fall within the range $[-1, 1]$. Further details of the RL environment and training process can be found in \cite{vanWijkAAS_23}. 

\begin{table}[htb]
    \centering
    \caption{NNC Expected Performance}
    \begin{tabular}{lcccc}
    \hline
        NNC                     & $\numpoints$      & $\deltav$         & Success   & Reward \\
    \hline
        baseline no sensors     & $98.6 \pm 0.49$   & $73.6 \pm 36.9$   & $59\%$    & $2.50 \pm 3.73$ \\
        best no sensors         & $95.3 \pm 13.7$   & $36.2 \pm 8.98$   & $93\%$    & $5.83 \pm 0.93$ \\
        baseline all sensors    & $90.5 \pm 13.9$   & $14.4 \pm 4.70$   & $40\%$    & $7.56 \pm 1.30$ \\
        best all sensors        & $96.5 \pm 10.8$   & $10.0 \pm 2.56$   & $78\%$    & $8.61 \pm 1.13$ \\
    \hline
    \end{tabular}
    \label{tab:nnc_metrics}
\end{table}

20 NNCs were trained for these experiments, ten for each of the variations described in \secref{sec:RL_setup}. From these, the four NNCs described in \tabref{tab:nnc_metrics} were selected for the experiments discussed in this paper. The first two are baseline NNCs with no sensors and all sensors, where the expected behavior of the agent is to simply orbit the chief spacecraft. The second two are the NNCs that achieved the highest reward for no sensors and all sensors, where the behavior of the agent is a more complex but more efficient maneuver.

\subsection{Lab Experiments}


Several experiments were conducted to demonstrate the NNC and RTA in the LINCS lab. Each experiment consisted of a single deputy agent with an initial state $[x, y, z, \dot{x}, \dot{y}, \dot{z}, \sunangle]=[21.8~\text{m}, -11.3~\text{m}, 41.8~\text{m}, 0~\text{m/s}, 0~\text{m/s}, 0~\text{m/s}, 3.42~\text{rad}]$. The experiments configurations are described in \tabref{tab:experiments}.
To ensure the trajectory of the spacecraft remains within the physical lab space and the experiments finish in a reasonable time, the simulated space frame is scaled into the lab frame. \tabref{tab:experiments} also shows the scale factors for each experiment, where space position/time are divided by the position/time scale to compute lab position/time. All observations and controllers are simulated in real time at a rate of 50 Hz, meaning that the NNC and RTA were run between 2.5 to 5 Hz, which is much faster than the RL environment's frequency of 0.1 Hz.

\begin{table}[htb]
    \centering
    \caption{Experiment Configurations}
    \label{tab:experiments}
    \begin{tabular}{c|c|c|c|c|c} \hline
        Experiment & Primary Controller & RTA & Illumination & Position Scale & Time Scale \\ \hline
        1 & NNC - no sensors & Off & Off & 65 & 10 \\
        2 & LQR & On & Off & 65 & 10 \\
        3 & NNC - no sensors & On & Off & 65 & 10 \\
        4 & NNC - all sensors & Off & On & 300 & 20 \\
        5 & Best NNC - no sensors & On & On & 65 & 15 \\
        6 & Best NNC - all sensors & On & On & 100 & 20 \\ \hline
    \end{tabular}
    
\end{table}

Experiments 1 and 3-6 use an NNC as the primary controller, where the objective is to complete the inspection task with or without RTA. Experiment 2 uses an LQR primary controller designed to push the system to the origin of Hill's frame and thus causes a collision. The objective of this experiment is to verify that the RTA can successfully intervene to avoid a collision.


The experiments are divided into two main categories: open loop and closed loop control. For open loop control, the NNC and RTA operate based on simulated space states for the entire simulation. The quadrotor is commanded to follow this simulated trajectory, but no information about the real world state is given to the NNC or RTA. This allows users to visualize the expected behavior of the NNC and RTA on a physical system. For closed loop control, the physical quadrotor state is converted to the space frame and provided to the NNC and RTA rather than the simulated space state. In this case, the state was determined using Vicon sensors. These experiments test the robustness of the NNC and RTA with noisy real world measurements and unexpected state data. While a quadrotor does not perfectly match a spacecraft operating in the real world, these experiments provide insight for how the NNC and RTA respond to a sim2real transfer. All 6 experiments were run with open loop control, and experiments 1 through 3 were repeated with closed loop control.

\section{Results}\label{sec:results}

For each experiment, the quadrotor position and velocity are recorded and compared to the simulated position and velocity. Additionally, the inspection points, thrust, and corresponding $\deltav$ are recorded based on the simulated position and control.

\subsection{Open Loop}

\begin{figure}
    \centering
    \includegraphics[width=\columnwidth]{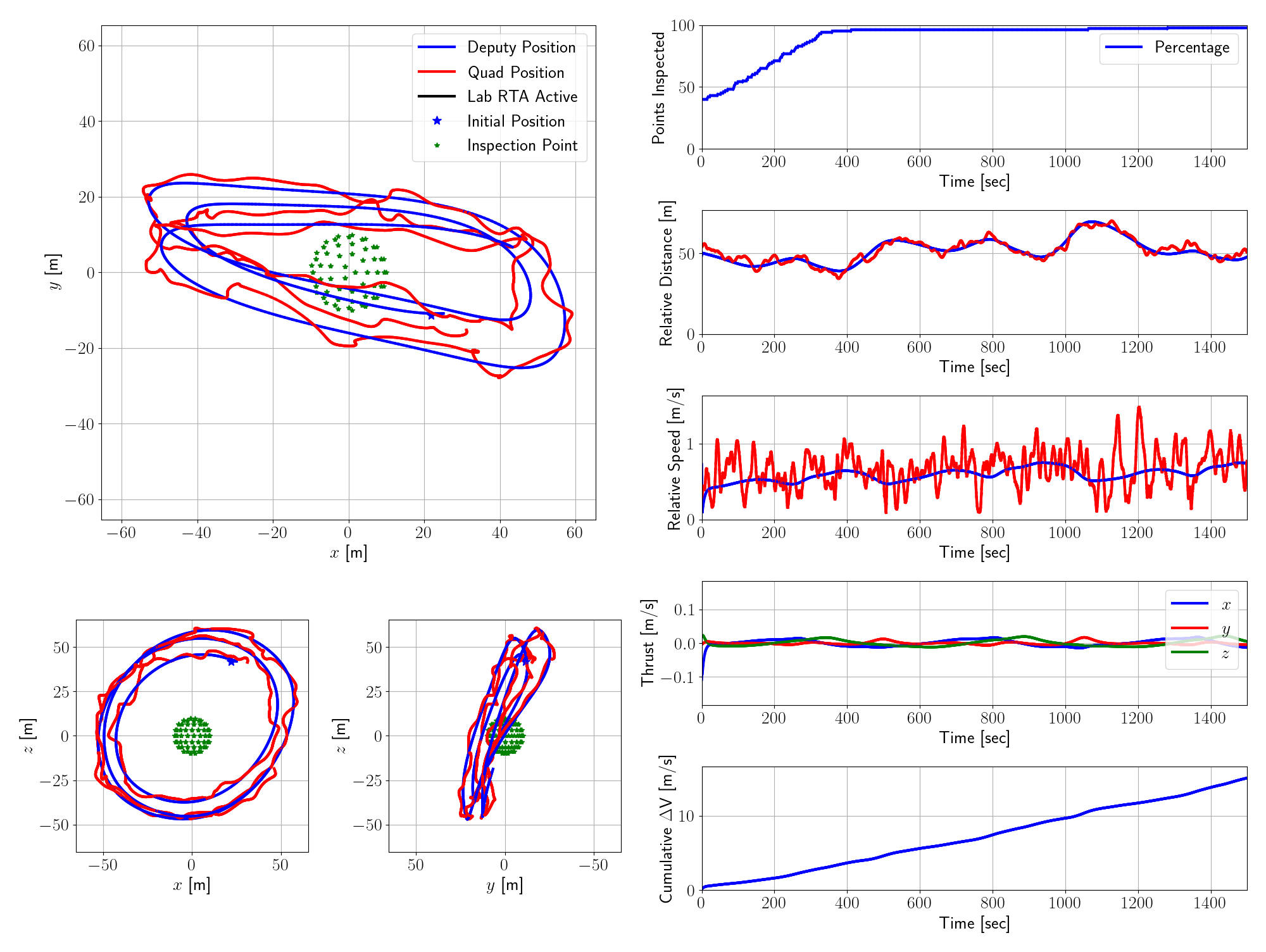}
    \caption{Simulated deputy state and sensed quadrotor state for experiment 1 (NNC - no sensors, RTA off, illumination off).}
    \label{fig:exp1}
\end{figure}

\begin{figure}
    \centering
    \includegraphics[width=.5\columnwidth]{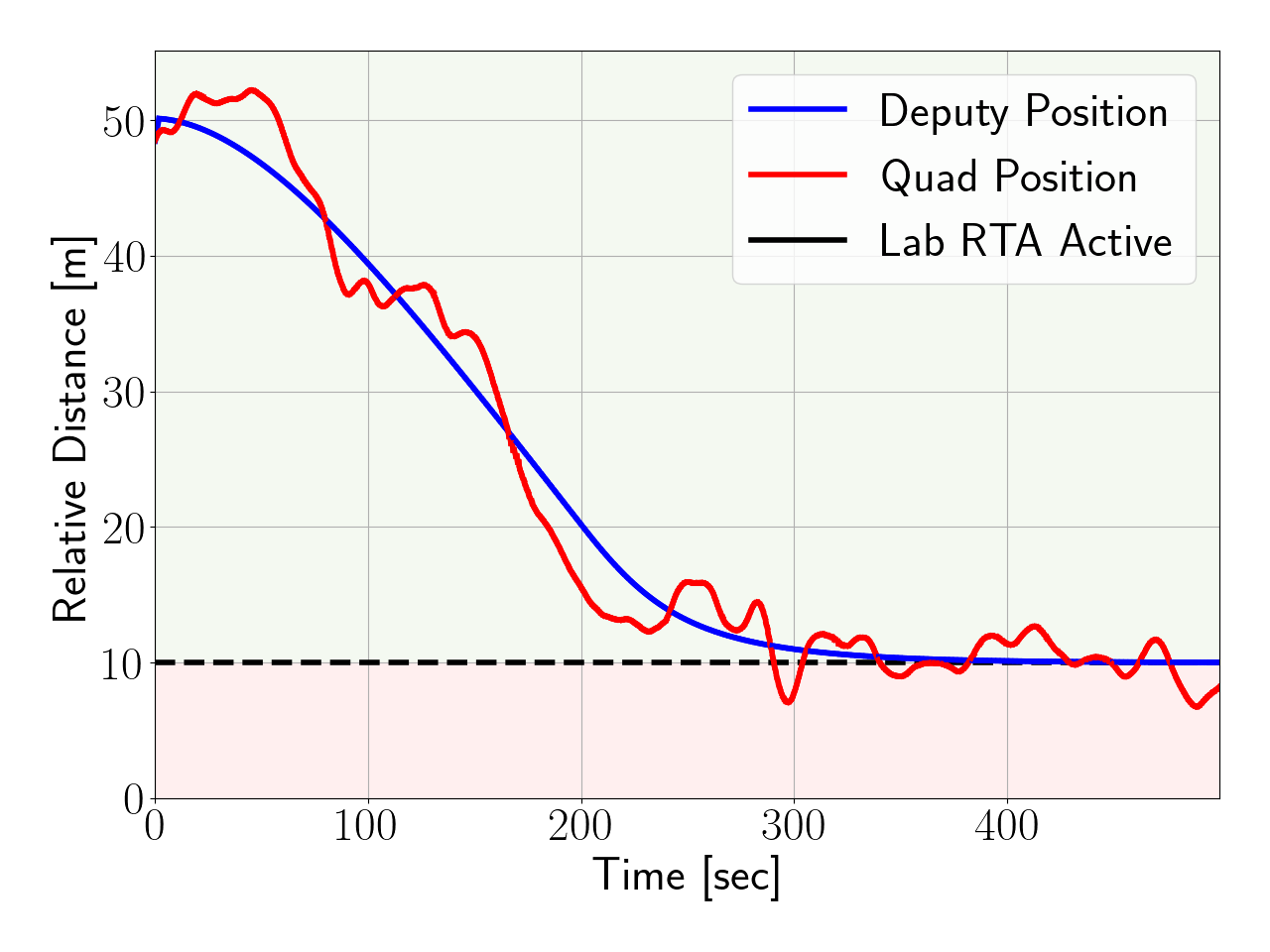}
    \caption{Simulated deputy relative distance to the chief and sensed quadrotor relative distance to the chief for experiment 2 (RTA only, open loop). The green shaded region is safe, the red shaded region is unsafe, and the black dotted line represents the collision radius.}
    \label{fig:exp2}
\end{figure}

\begin{figure}
    \centering
    \includegraphics[width=\columnwidth]{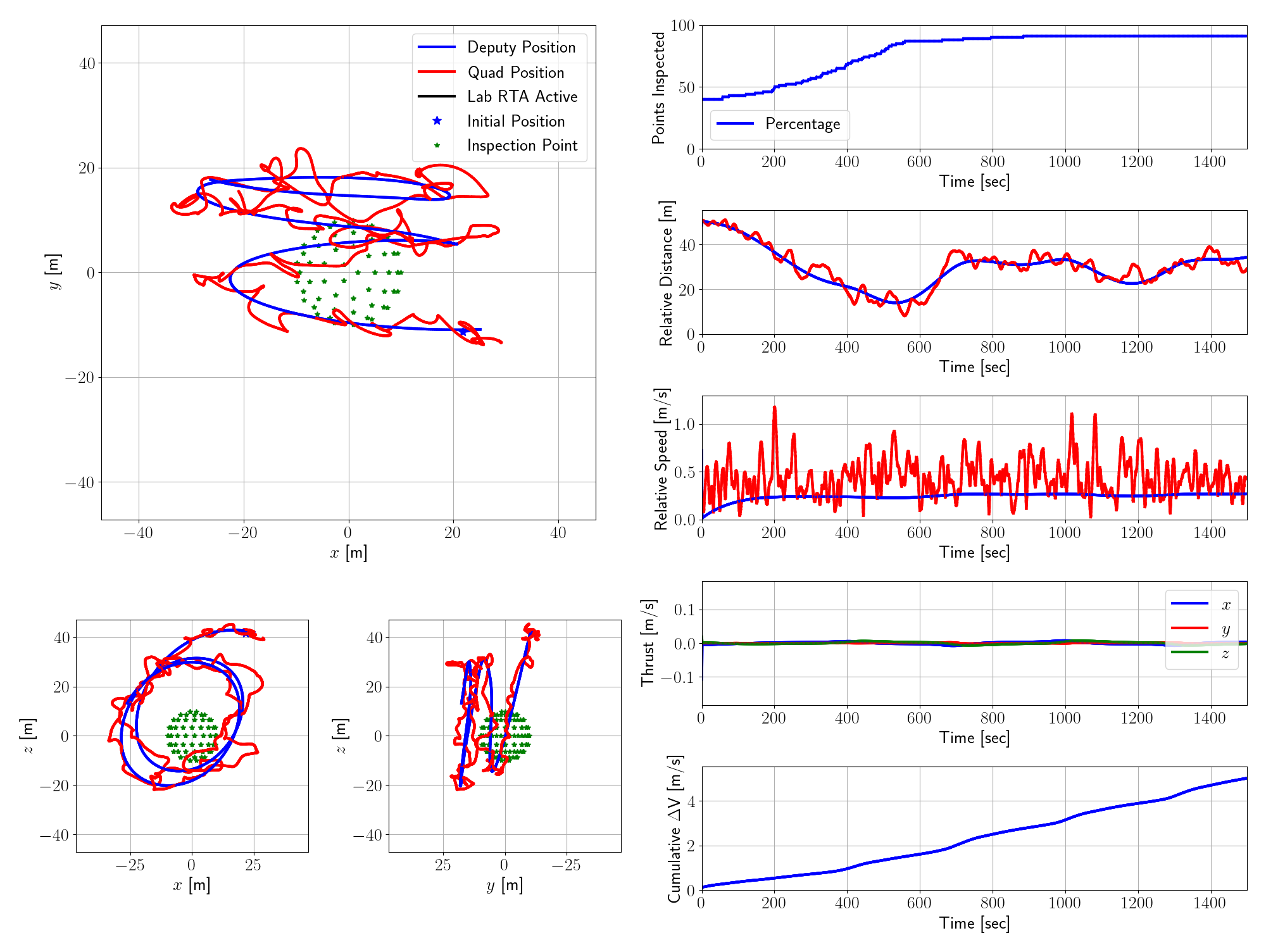}
    \caption{Simulated deputy state and sensed quadrotor state for experiment 3 (NNC - no sensors, RTA on, illumination off, open loop).}
    \label{fig:exp3}
\end{figure}

\begin{figure}
    \centering
    \includegraphics[width=\columnwidth]{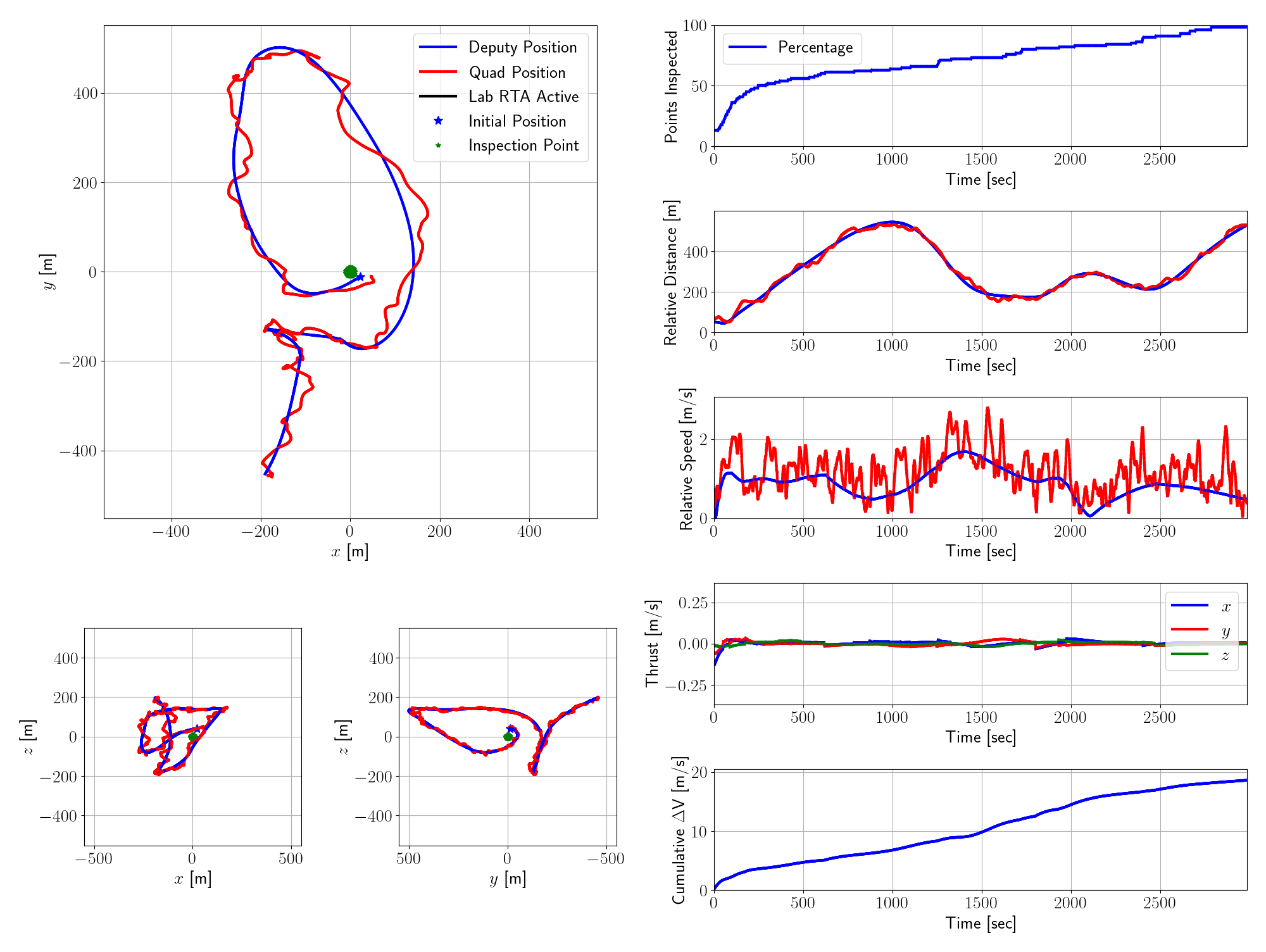}
    \caption{Simulated deputy state and sensed quadrotor state for experiment 4 (NNC - all sensors, RTA off, illumination on, open loop).}
    \label{fig:exp4}
\end{figure}

\begin{figure}
    \centering
    \includegraphics[width=\columnwidth]{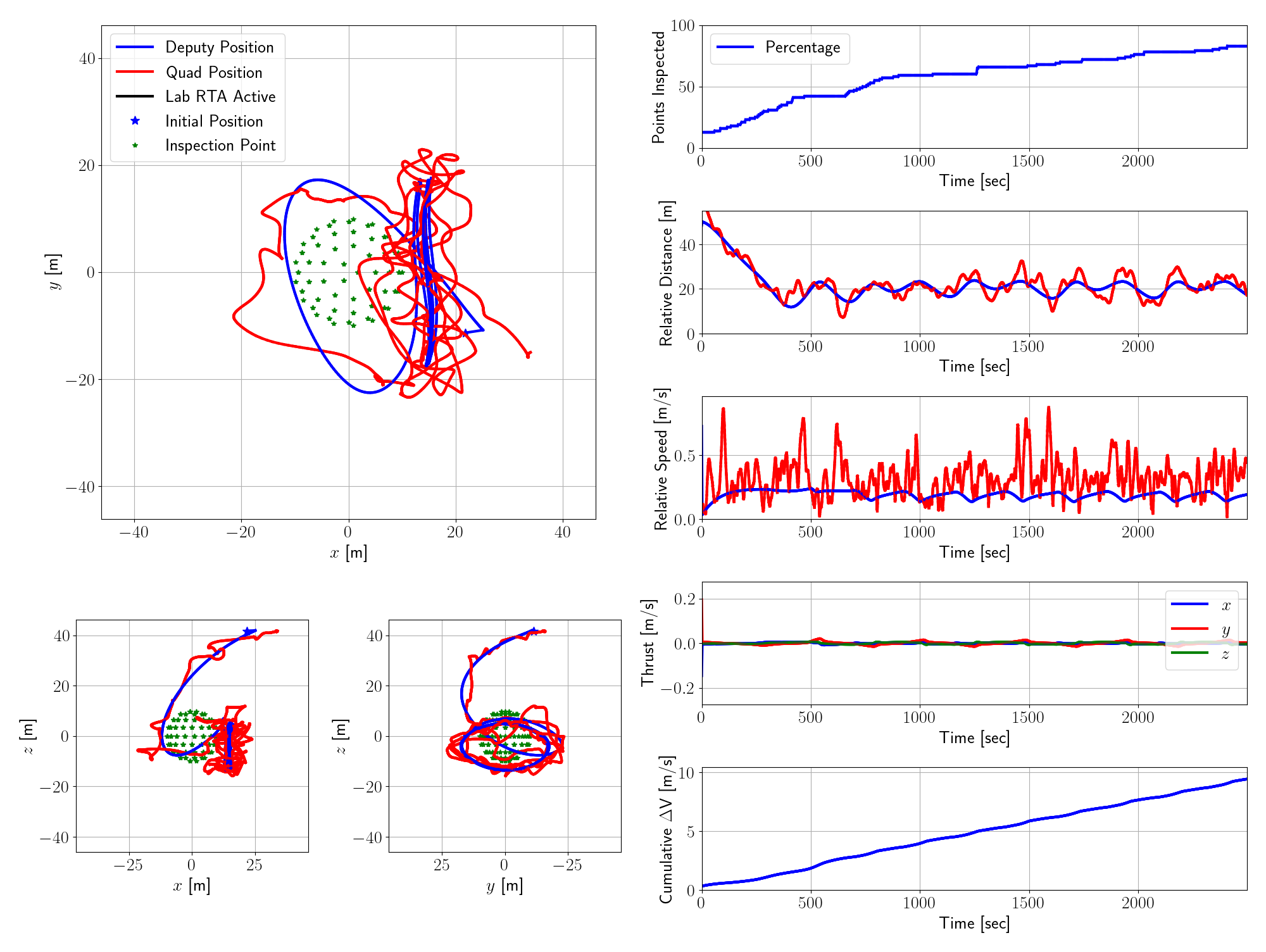}
    \caption{Simulated deputy state and sensed quadrotor state for experiment 5 (Best NNC - no sensors, RTA on, illumination on, open loop).}
    \label{fig:exp5}
\end{figure}

\begin{figure}
    \centering
    \includegraphics[width=\columnwidth]{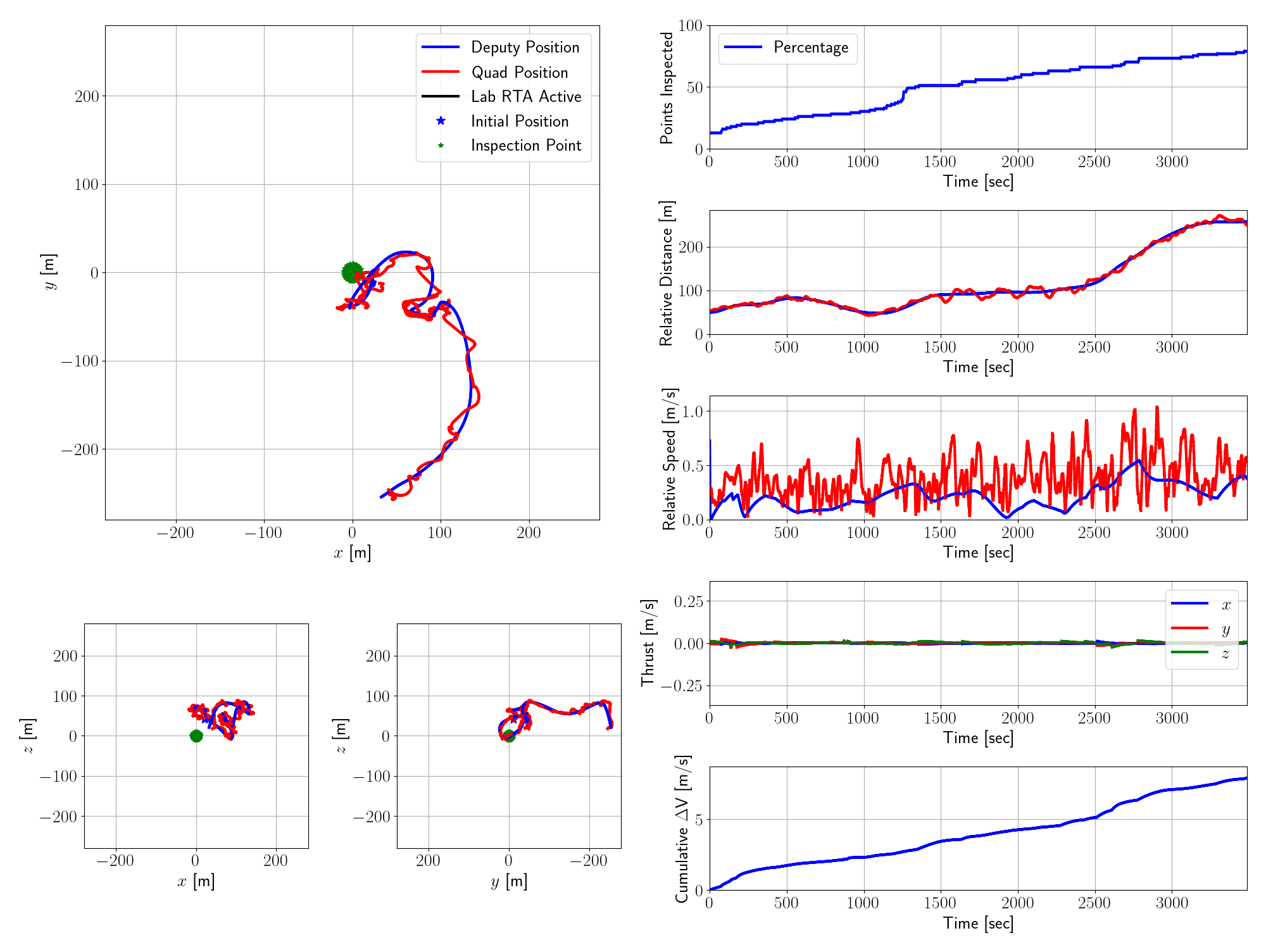}
    \caption{Simulated deputy state and sensed quadrotor state for experiment 6 (Best NNC - all sensors, RTA on, illumination on, open loop).}
    \label{fig:exp6}
\end{figure}

The 6 experiments conducted in open loop simulation are presented in \figref{fig:exp1} through \figref{fig:exp6}. First, \figref{fig:exp1} shows the trajectory of the deputy for \textbf{experiment 1} along with the simulated inspected points, thrust, and $\deltav$. The trajectory shows this agent learns to complete a near circular orbit in the $x-z$ plane in order to circumnavigate the chief and inspect all points. Without illumination, the agent inspects almost all of the points within 350 seconds using $<5$~m/s of $\deltav$, but it keeps orbiting and using more $\deltav$ to inspect the last few points. When comparing the trajectories of the simulated deputy and sensed quadrotor, the quadrotor generally follows the simulated trajectory with some minor deviations due to real world disturbances. 
For this experiment, it is important to note that for every meter that the quadrotor deviates from the simulated position in the space frame, this corresponds to only a 1.5~cm deviation in the real world.
The relative speed of the quadrotor appears to be a much noisier signal when compared to the simulated relative speed. Overall, the tracking capability of the quadrotor allows users to visualize the approximate trajectory of the agent.

\figref{fig:exp2} shows the relative distance over time for the simulated deputy and sensed quadrotor for \textbf{experiment 2}. The trajectory shows how the LQR primary controller pushes the deputy towards a collision, but the RTA intervenes and keeps the deputy at a safe distance of $10$~m away. The quadrotor again generally follows the simulated trajectory, but does enter the unsafe set when the simulate trajectory nears the boundary. This is not unexpected, as the RTA has no knowledge of the quadrotor's state, and has no reason to further modify the control. While methods such as robust CBFs \cite{jankovic2018robust} exist that account for disturbances, this experiment shows that true state information is important when using RTA to assure safety.

\figref{fig:exp3} shows the trajectory of the deputy for \textbf{experiment 3}, where RTA is applied to the same NNC as experiment 1. In this experiment, the RTA frequently intervenes to enforce speed limits, which caused the agent to orbit the chief at a slightly smaller relative distance. As a result, this intervention prevents the agent from inspecting all points, but it also uses $10$~m/s less $\deltav$. This experiment shows that RTA can successfully assure safety for an NNC that was not trained with safety in mind, while allowing the NNC to stay on task as much as possible.

\figref{fig:exp4} shows the trajectory of the deputy for \textbf{experiment 4}, where the NNC is trained with all sensors. The results show the deputy completing a much larger, slower orbit that attempts to follow the Sun around the inspection points. \figref{fig:exp5} and \figref{fig:exp6} show the trajectories for \textbf{experiments 5 and 6} respectively, where the best performing NNC for no sensors or all sensors was used with RTA. These results are more complex trajectories, where the NNC has learned to maximize reward in the environment using the available observations. In all cases, the quadrotor generally follows the simulated trajectory with minor disturbances, and provides a simple visualization to help understand how the agents would perform in the real world.

\subsection{Closed Loop}

\begin{figure}
    \centering
    \includegraphics[width=\columnwidth]{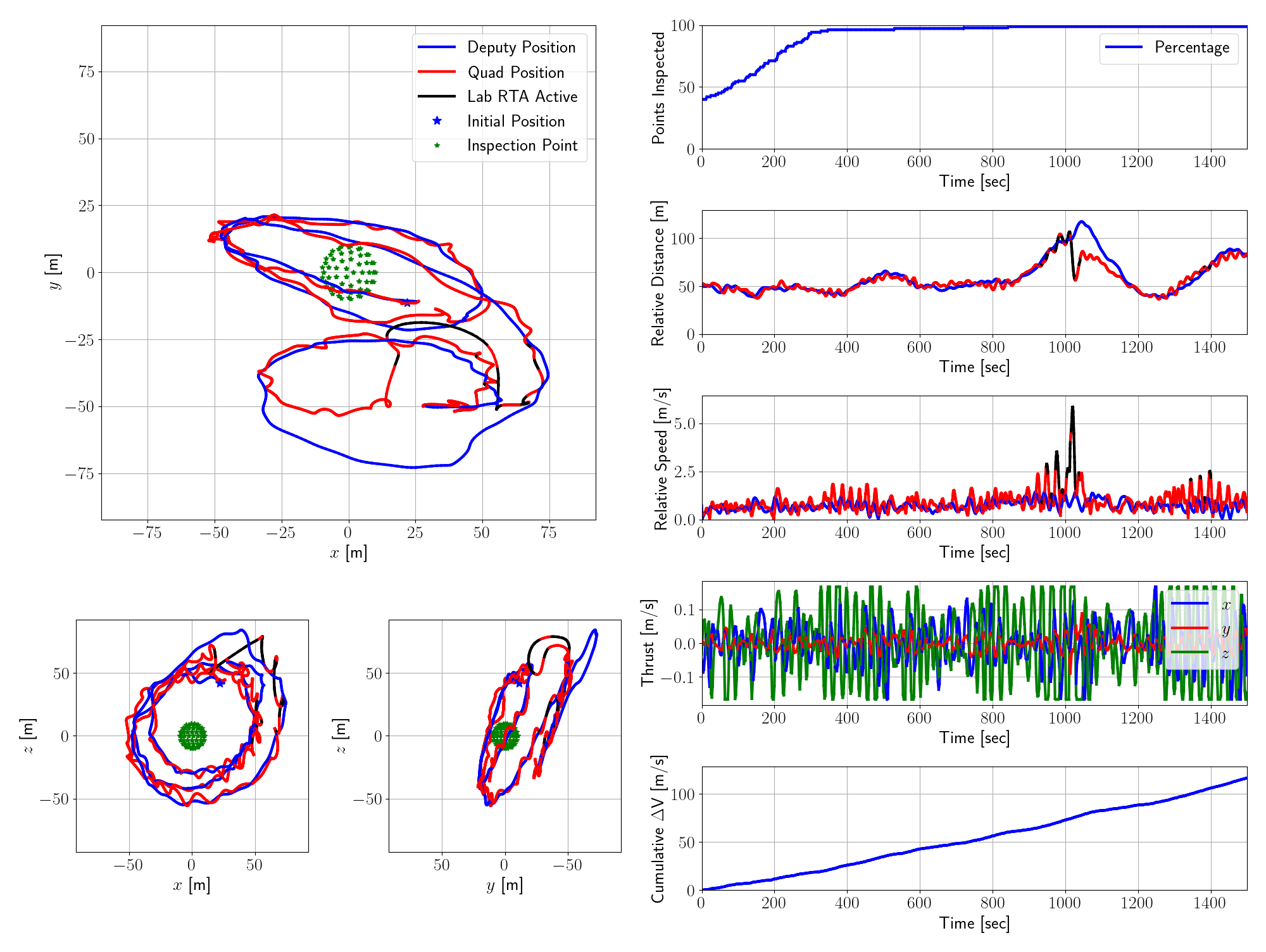}
    \caption{Simulated deputy state and sensed quadrotor state for experiment 1 (NNC - no sensors, RTA off, illumination off, closed loop), where the sensed state is passed to the NNC.}
    \label{fig:exp1_vicon}
\end{figure}

For the closed loop simulations, experiments 1 through 3 were repeated where Vicon sensors were used to feed the quadrotor state back to the NNC and RTA.
Since the noisy quadrotor state data is fed back to the NNC and RTA, it is expected that the trajectories will slightly differ from the open loop simulations.
\figref{fig:exp1_vicon} shows the trajectory of the deputy for \textbf{experiment 1}. From the start, the trajectory is much noisier than the open loop simulation in \figref{fig:exp1}. The thrust specifically shows considerable chattering, where the agent is constantly attempting to make adjustments to the trajectory based on the noisy state inputs. After deputy has completed one orbit around the chief, the trajectory begins to deviate from the expected path, and pushes the quadrotor towards the physical wall of the lab. As a result, near the position $[x, y] = [50, -50]$~m, the lab RTA onboard the quadrotor intervenes to prevent a collision with the lab wall, causing a large deviation between the deputy position and quadrotor position. Once the deputy returns to a safe state, the quadrotor begins to track this trajectory again. To avoid this scenario, the position scaling for the experiment could have been increased, such that the simulated trajectory does not get as close to the lab wall. Overall, this experiment shows that the NNC can adapt to noisy observations and states that it does not expect, but this can alter the expected behavior of the agent.

\begin{figure}
    \centering
    \includegraphics[width=.5\columnwidth]{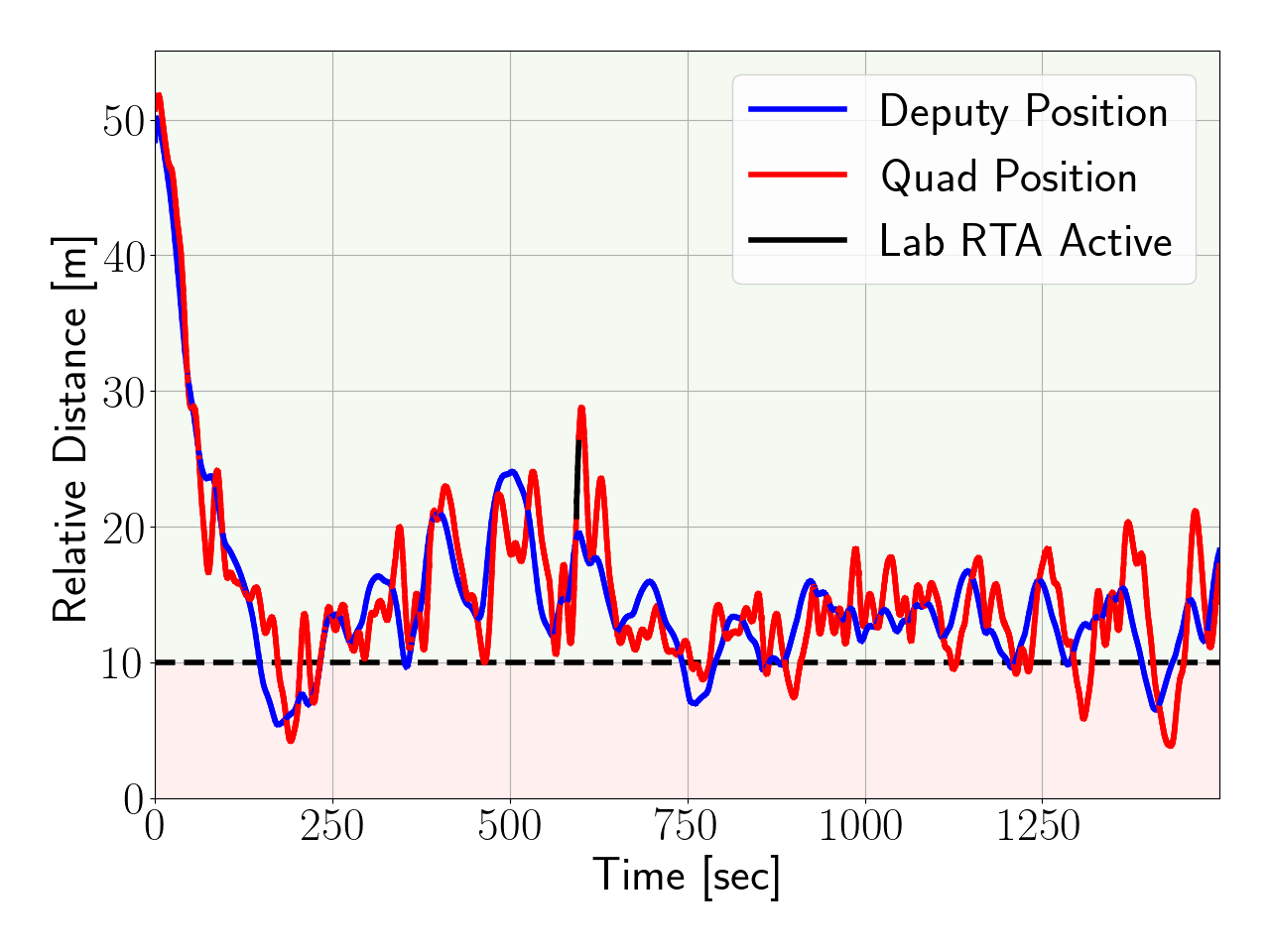}
    \caption{Simulated deputy relative distance to the chief and sensed quadrotor relative distance to the chief for experiment 2 (RTA only, closed loop), where the sensed state is passed to the RTA. The green shaded region is safe, the red shaded region is unsafe, and the black dotted line represents the collision radius.}
    \label{fig:exp2_vicon}
\end{figure}

\figref{fig:exp2_vicon} shows the relative distance over time for the simulated deputy and sensed quadrotor for \textbf{experiment 2} for the closed loop experiment. In this case, the results show both the simulated and sensed positions enter the unsafe set. Importantly, due to the properties of a zeroing CBF, the deputy stabilizes around the boundary of $10$~m and prevents a complete failure where the LQR primary controller would push the deputy to the origin. Again, a technique such as robust CBFs could be applied to prevent any safety violations.

\begin{figure}
    \centering
    \includegraphics[width=\columnwidth]{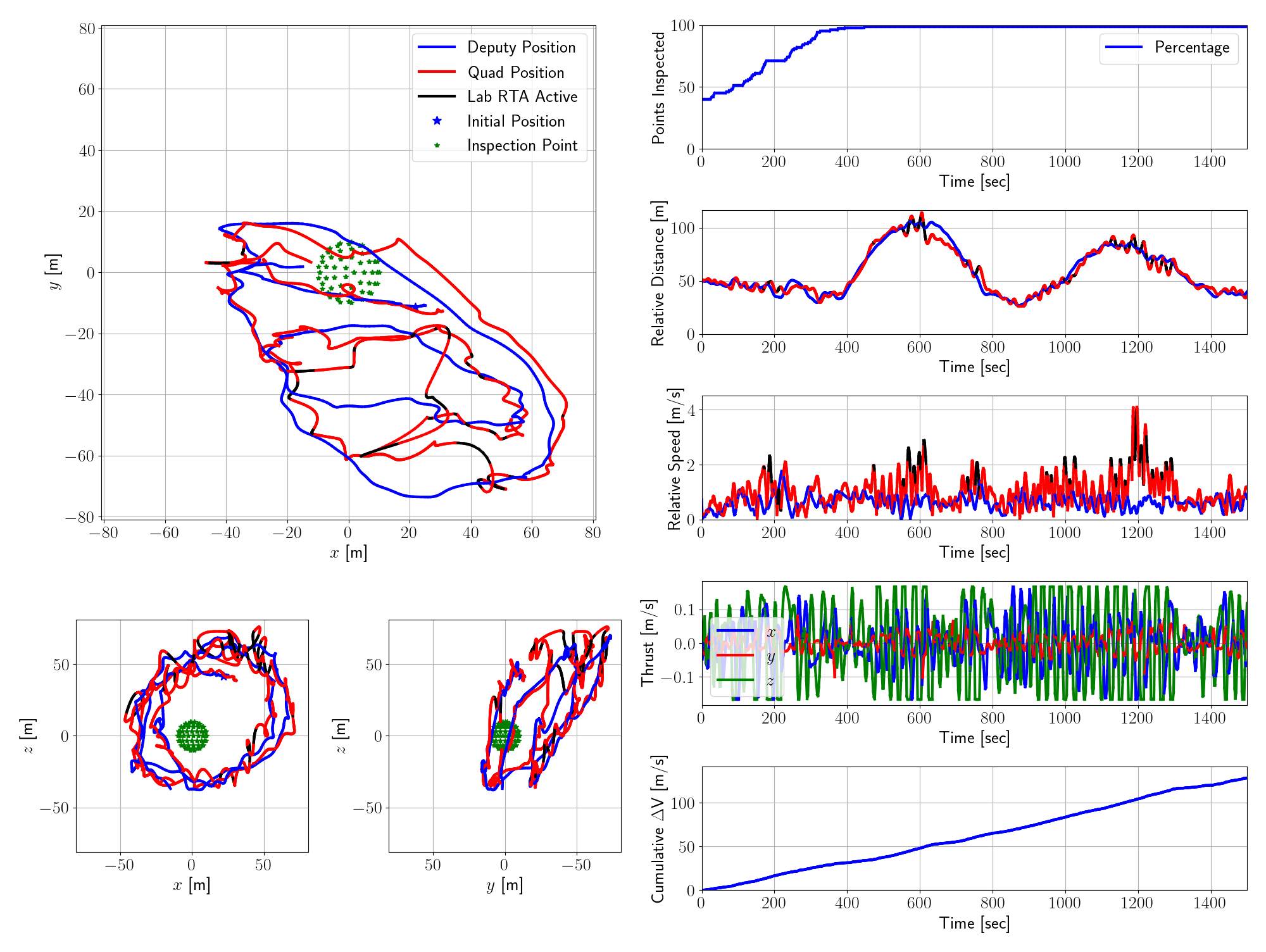}
    \caption{Simulated deputy state and sensed quadrotor state for experiment 3 (NNC - no sensors, RTA on, illumination off, closed loop), where the sensed state is passed to the NNC and RTA.}
    \label{fig:exp3_vicon}
\end{figure}

Finally, \figref{fig:exp3_vicon} shows the trajectory of the deputy for \textbf{experiment 3} with both the NNC and RTA. The results are very similar to experiment 1, where the noisy observations cause the simulated deputy to move closer to the lab wall, where the lab RTA must intervene to prevent a collision. Overall, these results demonstrate that the NNC and RTA can work together in the presence of disturbances and noise.

\section{Conclusion} \label{sec:conclusion}

In this paper, several different experimental results were presented showing an NNC and RTA filter simulated in the LINCS lab. First, the experiments were run in open loop with only simulated data passed to the NNC and RTA, where the behavior of the control system can be visualized on a real world platform. The quadrotors were shown to track the simulated space trajectories with minimal deviation. Second, the experiments were run in closed loop with real world data passed to the NNC and RTA, which tests the robustness of the control system to real world disturbances. The NNC and RTA were shown to be robust to these disturbances where the agent can continue the desired task, and minor improvements to the RTA will lead to assuring safety in all scenarios. Future work should consider minimizing the effect of these disturbances and testing more complex control systems in the LINCS lab, such as multiagent scenarios or six degree-of-freedom agents.



\section*{Acknowledgments}
This research was sponsored by the Air Force Research Laboratory under the Safe Trusted Autonomy for Responsible Spacecraft (STARS) Seedlings for Disruptive Capabilities Program.
The views expressed are those of the authors and do not reflect the official guidance or position of the United States Government, the Department of Defense, or of the United States Air Force. This work has been approved for public release: distribution unlimited. Case Number AFRL-2024-2551.

\bibliography{references}

\end{document}